\documentclass[conference]{IEEEtran}
\IEEEoverridecommandlockouts

\usepackage{cite}
\usepackage{amsmath,amssymb,amsfonts}
\usepackage{graphicx}
\usepackage{textcomp}
\usepackage{xcolor}
\usepackage{booktabs}
\usepackage{pifont}
\usepackage{mathtools}
\usepackage{caption}       
\usepackage[linesnumbered,ruled,vlined]{algorithm2e}
\usepackage{url} 
\usepackage{tabularx} 
\usepackage{multirow} 
\usepackage{hyperref}
\usepackage{graphicx}
\usepackage{subcaption} 
\usepackage{pdfrender}

\def\BibTeX{{\rm B\kern-.05em{\sc i\kern-.025em b}\kern-.08em
    T\kern-.1667em\lower.7ex\hbox{E}\kern-.125emX}}
\begin{document}

\newcommand*{\boldcheckmark}{%
  \textpdfrender{
    TextRenderingMode=FillStroke,
    LineWidth=.3pt, 
  }{\checkmark}%
}
\newcommand{\cmark}{\textcolor{green!70!black}{\boldcheckmark}}%
\newcommand{\xmark}{×}

\SetCommentSty{mycommfont}
\newcommand{\mycommfont}[1]{\textcolor{green!50!black}{\texttt{#1}}}

\let\oldemptyset\emptyset
\let\emptyset\varnothing

\title{FlexCTC: GPU-powered CTC Beam Decoding With Advanced Contextual Abilities}

\author{
\IEEEauthorblockN{
Lilit Grigoryan\IEEEauthorrefmark{1}\IEEEauthorrefmark{3}\IEEEauthorrefmark{4}, 
Vladimir Bataev\IEEEauthorrefmark{1}\IEEEauthorrefmark{3}, 
Nikolay Karpov\IEEEauthorrefmark{1}, 
Andrei Andrusenko\IEEEauthorrefmark{1},
Vitaly Lavrukhin\IEEEauthorrefmark{2}
and Boris Ginsburg\IEEEauthorrefmark{2}}
\newline
\IEEEauthorblockA{\IEEEauthorrefmark{1}\textit{NVIDIA}, Yerevan, Armenia}
\IEEEauthorblockA{\IEEEauthorrefmark{2}\textit{NVIDIA}, Santa Clara, USA}
\IEEEauthorblockA{\IEEEauthorrefmark{3}Equal contribution}
\IEEEauthorblockA{\IEEEauthorrefmark{4}Corresponding author, \textit{lgrigoryan@nvidia.com}}
}


\maketitle

\begin{abstract}
While beam search improves speech recognition quality over greedy decoding, standard implementations are slow, often sequential, and CPU-bound. To fully leverage modern hardware capabilities, we present a novel open-source FlexCTC toolkit for fully GPU-based beam decoding, designed for Connectionist Temporal Classification (CTC) models. Developed entirely in Python and PyTorch, it offers a fast, user-friendly, and extensible alternative to traditional C++, CUDA, or WFST-based decoders. The toolkit features a high-performance, fully batched GPU implementation with eliminated CPU-GPU synchronization and minimized kernel launch overhead via CUDA Graphs. It also supports advanced contextualization techniques, including GPU-powered N-gram language model fusion and phrase-level boosting. These features enable accurate and efficient decoding, making them suitable for both research and production use.
\end{abstract}


\section{Introduction}

Advancements in GPU hardware and deep learning architectures have facilitated the full parallelization of many components in automatic speech recognition (ASR) systems on GPUs. Modern ASR encoder architectures -- such as transformers \cite{ article:transformer2,article:transformer} and Conformers \cite{article:conformer} -- are explicitly engineered to leverage this parallelism by enabling simultaneous computation across audio sequences, thereby maximizing GPU utilization and throughput. In contrast, the decoding phase -- particularly beam search is often executed on the CPU due to the limited availability of flexible GPU-oriented implementations. This work presents a novel batched beam decoding approach optimized specifically for efficient execution on GPUs. We will focus on Connectionist Temporal Classification (CTC) \cite{article:ctc} models.


Beam search is inherently sequential, as each step depends on the hypotheses generated in the previous one, which makes GPU parallelization limited and more challenging. However, beam search can still benefit significantly from batching and parallelism. Batching refers to processing multiple utterances in parallel, rather than decoding them one by one. In addition to this input -- level batching, we can also exploit intrainput parallelism by processing multiple hypotheses in beam in parallel. By combining these two levels of parallelism -- across inputs and across beam hypotheses per input -- and executing them on the GPU, we can substantially accelerate the decoding process.
Additionally, beam search performance is closely tied to beam size: larger beams improve the chance of finding high-scoring hypotheses but increase computational cost. Our implementation exploits intra-input parallelism on GPUs, enabling high-throughput decoding even with large beam widths.

Beam search decoding becomes especially effective when enhanced with customization techniques that guide the recognition process toward more accurate and relevant results. One common method is language model (LM) fusion, which combines an external LM with the acoustic model’s output scores to improve decoding. The external LM provides linguistic knowledge learned from large text datasets, helping to enhance the acoustic model’s predictions with better syntax and semantics. There are several ways to integrate the LM, such as shallow fusion~\cite{article:shallowfusion2015}, cold fusion, and deep fusion, which differ in how the LM influences the decoding process~\cite{article:fusioncomparison}. In addition to this general guidance, phrase-level boosting provides a targeted approach to increase the likelihood of specific user-defined phrases, such as personal names, technical terms, or brand names.
We integrate a GPU-accelerated n-gram language model (NGPU-LM) for both LM fusion and word boosting, making the entire decoding pipeline fully GPU-based.

In this work, we introduce a novel set of tools (i.e., toolkit) that is required for fully GPU-based beam decoding designed for CTC models. FlexCTC\footnote{https://github.com/NVIDIA/NeMo/pull/13337} features the following key advancements:

\begin{itemize}
    \item \textbf{High-speed performance:} At the core of the toolkit is a GPU-optimized, fully batched implementation of beam decoding for CTC models, with minimal CPU-GPU interaction overhead optimized through CUDA Graphs integration.
    \item \textbf{Contextualization:} The toolkit supports advanced contextualization techniques such as N-gram LM shallow fusion and phrase boosting, powered by the GPU-native NGPU-LM module. These features offer a significant accuracy boost, with minimal runtime overhead.
    \item \textbf{Flexibility:} Unlike other CTC decoders implemented in C++, CUDA or using WFST (Weighted Finite-State Transducer) graphs, our toolkit is developed entirely in Python and PyTorch, offering ease of use and flexibility for both research and production purposes.
\end{itemize}
The full toolkit is released as open source under a production-friendly license in NeMo framework \cite{article:nemo} to support real-world deployment and accelerate further research and innovation.

\section{Related Work}
\subsection{CTC models}
CTC is a widely adopted framework for sequence-to-sequence learning in modern end-to-end ASR systems. 
Its non-autoregressive nature allows tokens to be emitted independently at each time step, enabling significantly faster inference compared to autoregressive approaches like Transducers and Attention-based Encoder-Decoders (e.g., Transformers). 
Thanks to its efficiency and simplicity, CTC is also a popular choice in production environments.

Unlike Transducers \cite{article:rnnt}, which maintain hidden states, or attention-based models that rely on cached representations, CTC models operate without internal state or memory and are less likely to memorize domain-specific patterns from the training data. This makes them especially easy to adapt to new domains during decoding using techniques such as n-gram LM fusion or phrase-level boosting.  These customization methods are more effective during beam search decoding, where multiple hypotheses are tracked and rescored, allowing the customization to boost scores of domain-specific words or phrases and increase their chances of being selected as the final output.

\subsection{CTC beam decoders}
Several CPU-based CTC beam search decoders are publicly available, with Flashlight~\cite{article:flashlight} being among the most widely used. Flashlight is a highly optimized C++ framework that implements CTC beam search with support for KenLM-based~\cite{article:kenlm} language model fusion. However, it does not support word boosting. It also supports streaming inference. 
PyCTCDecode~\cite{article:pyctcdecode} is a Python-based library that provides a CPU-focused CTC beam search decoder. It supports word boosting and n-gram language model fusion. 

GPU-based ASR decoding methods fall into two categories: WFST-based and non-WFST approaches. WFST-based systems compile acoustic, lexical, and language model components into a single decoding graph. This enables efficient deterministic search, but limits runtime flexibility and can be memory-intensive. In \cite{article:galvez} open-source GPU-accelerated WFST decoder is introduced. It supports online inference, n-gram LM fusion and utterance-specific word boosting via on-the-fly composition. CUDA Graphs are also used to eliminate GPU kernel launch overhead. 

Non-WFST decoders, in contrast, implement search logic programmatically and offer more flexibility for integrating neural LMs and biasing mechanisms. They are also better suited for fine-grained GPU optimizations. Seki et al.~\cite{article:vec_beam_search} proposed a vectorized beam search algorithm for joint CTC/attention decoding with Recurrent Neural Network LM. Their method processes multiple beam hypotheses in parallel - intra-input parallelism introduced earlier. However, the implementation was not made publicly available. In contrast, PyTorch’s CUCTCDECODER \cite{article:cuctcdecoder} is an open-source GPU-accelerated CTC decoder implemented in CUDA. For speed up with minimal loss in accuracy it uses \texttt{blank\_skip\_threshold} parameter, which skips frames with high probability of blank symbol. Unfortunately, the current implementation does not support LM fusion and word boosting. 

This work introduces a non-WFST GPU-accelerated CTC beam decoding method that supports language model fusion and word boosting. Summarization of existing implementations is shown in Table \ref{tab:ctc_decoder_comparison}.


\begin{table}[t]
    \centering
    \begin{tabular}{l@{\hspace{6pt}}cccccc}
        \toprule
         & \textbf{Open} & \textbf{Python} & \textbf{CPU} & \textbf{GPU} & \textbf{LM} & \textbf{PB} \\
        \midrule
        \cite{article:flashlight} Flashlight             & \cmark & \xmark & \cmark & \xmark & \cmark & \xmark \\
        \cite{article:pyctcdecode} PyCTCDecode           & \cmark & \cmark & \cmark & \xmark & \cmark & \cmark \\
        \cite{article:galvez} Cuda WFST Dec.             & \cmark & \xmark & \xmark & \cmark & \cmark & \cmark \\
        \cite{article:vec_beam_search} Vec. Beam Dec.    & \xmark & \xmark & \xmark & \cmark & \cmark & \xmark \\
        \cite{article:cuctcdecoder} CUCTCDecoder         & \cmark & \xmark & \xmark & \cmark & \xmark & \xmark \\
        \midrule
        \textbf{Our: FlexCTC}                            & \cmark & \cmark & \cmark & \cmark & \cmark & \cmark \\
        \midrule
    \end{tabular}
    \caption{Comparison of features in existing CTC beam search decoders. \textbf{LM}: LM fusion, \textbf{PB}: phrase boosting}
    \label{tab:ctc_decoder_comparison}
\end{table}

\subsection{Accelerating Batched Beam Search with Trie}
Our algorithmic optimizations are inspired by~\cite{article:rnnt_batched_decoding}, which proposes GPU-efficient techniques for beam search decoding in Transducer-based models. This includes a trie-like structure for managing hypotheses and a hashing-based strategy for fast comparison and merging. These design choices enable batched execution, improving GPU utilization and reducing the decoding complexity from \( O(N^2) \) to \( O(N) \) with respect to the audio sequence length. Additionally, the use of CUDA Graphs helps eliminate GPU kernel launch overhead, further enhancing decoding throughput.

\subsection{Statistical Language Models: NGPU-LM}
Context biasing in ASR systems is often performed using statistical language models. One common technique is shallow fusion, where probability distributions from the ASR model and an external LM are combined during the decoding step.

A widely adopted choice for n-gram LMs is KenLM~\cite{article:kenlm}, a fast and efficient CPU-based toolkit. However, integrating KenLM into GPU-based decoding pipelines can introduce latency due to CPU-GPU synchronization overhead, which reduces the benefits of GPU parallelization.

To address this limitation, NGPU-LM~\cite{article:ngpulm} has been proposed, an n-gram language model specifically optimized for GPU parallelism. NGPU-LM supports fully GPU-based execution and enables batched query processing, significantly improving throughput. In~\cite{article:ngpulm}, NGPU-LM is used for external language model fusion during greedy decoding, while~\cite{article:ngpulm_wb} presents GPU-PB, a phrase boosting method based on NGPU-LM and demonstrates its use during greedy and beam search decoding. By replacing traditional CPU-based LMs with NGPU-LM, LM fusion and word boosting can be performed entirely on the GPU, eliminating CPU-side computations and enabling fully GPU-based context biasing.

\section{Method}

\subsection{Hypotheses organization}
We adopt the \texttt{BatchedBeamHyps} data structure introduced in~\cite{article:rnnt}, which combines a trie-like organization of hypotheses with efficient hash-based comparisons. The trie structure allows hypotheses to share common prefixes, improving memory efficiency and enabling fast expansion without duplicating full transcripts. Transcripts are stored using token and pointer tensors, which allows for reconstruction of complete sequences on demand. To enable efficient hypothesis merging, each hypothesis maintains an incremental hash value that is updated only when a new token is appended. In the case of CTC models, the hash is updated only for non-blank and non-repeated tokens, in accordance with the CTC decoding rule. 

\subsection{LM fusion and word-boosting}
For fast LM fusion we use NGPU-LM, which enables efficient querying over the entire vocabulary. Phrase boosting is handled by GPU-PB, which combines a phrase prefix tree—constructed using the Aho-Corasick algorithm with NGPU-LM’s GPU-optimized architecture. Unlike approaches that apply rewards only at phrase endpoints, GPU-PB progressively distributes boosting scores along the prefix tree based on node depth.

The overall decoding score for a hypothesis is computed as:
\begin{equation}
    \text{s} = \log P_{\mathrm{CTC}} + \alpha_{\mathrm{LM}} \log P_{\mathrm{LM}} + \alpha_{\mathrm{BT}} \log P_{\mathrm{PB}} + \beta \cdot N
\end{equation}

where \(P_{\mathrm{CTC}}\) is the probability from CTC, \(P_{\mathrm{LM}}\) is the LM probability from NGPU-LM, \(P_{\mathrm{PB}}\) is the phrase boosting score from GPU-PB, \(\alpha_{\mathrm{LM}}\) and \(\alpha_{\mathrm{BT}}\) are fusion weights for the LM and phrase boosting respectively, \(\beta\) is the insertion penalty weight, and \(N\) is the number of tokens in the hypothesis.

\begin{algorithm}[t]
\LinesNotNumbered

\footnotesize
\caption{CTC Beam Search Decoding}


\SetKwInOut{Input}{Input}
\SetKwInOut{Output}{Output}
\Input{Log probs $D \in \mathbb{R}^{B \times T \times |V|}$}
\Input{Lengths $L \in \mathbb{N}^{B}$ \hfill \tcp{Valid sequence lengths}}
\Output{\texttt{beams} -- batched hypotheses in trie-like structures}

\hfill \\

\tcp{Initialize accumulated scores $\in \mathbb{R}^{B \times K}$}
$\texttt{acc\_scores[:,\, 0]} = 0,\quad \text{else } -\infty$\;
\If{$\texttt{LM}$}{
    \tcp{Initialize with SOS scores from LM}
    $\texttt{lm\_scores},\,\texttt{lm\_sts} \gets \mathrm{LM}\left(\texttt{<SOS>}\right)$
}
\If{$\texttt{BT}$}{
    $\texttt{bt\_scores},\,\texttt{bt\_sts} \gets \mathrm{BT}\left(\texttt{<0>}\right)$
}
\For{$t \gets 0$ \KwTo $T - 1$}{
    $\texttt{active\_mask} \gets \left(t < L\right)$ \;
    $\texttt{blanks\_mask} \gets (\texttt{blank\_label} == V)$ \;
    $\texttt{repeat\_mask} \gets (\texttt{beams.last\_labels} == V)$ \;
    $\texttt{rb\_mask} \gets (\texttt{repeat\_mask} \lor \texttt{blanks\_mask})$ \;
    \hfill \\
    \tcp{Update scores logp $\in \mathbb{R}^{B \times K \times |V|}$}
    $\texttt{logp} \gets D[:\,, t\,, :] + \texttt{acc\_scores}$\;
    $\texttt{logp} \mathrel{+}= \beta \cdot (1 - \texttt{rb\_mask})$\;

    \If{$\texttt{LM}$}{
        $\texttt{logp}[:, :, \neg \texttt{rb\_mask}] \mathrel{+}= \alpha_{LM} \cdot \texttt{lm\_scores}$ \;
    }
    \If{$\texttt{BT}$}{
        $\texttt{logp}[:, :, \neg \texttt{rb\_mask}] \mathrel{+}= \alpha_{BT} \cdot \texttt{bt\_scores}$ \;
    }

    \hfill \\
    \tcp{Select labels: \texttt{logp} $\in \mathbb{R}^{B \times \left(K * |V|\right)}$}
    $\texttt{acc\_scores}, \texttt{indices} \gets \mathrm{TopK}\left(\texttt{logp}, \mathrm{K}\right)$ \;
    $\texttt{new\_labels} \gets \texttt{indices} \bmod V$ \;
    $\texttt{new\_beamid} \gets \left\lfloor \texttt{indices} / V \right\rfloor$ \;

    \hfill \\
    $\texttt{max\_score} \gets \max(\texttt{acc\_scores})$ \;
    $\texttt{acc\_scores}(\texttt{acc\_scores} < \texttt{max\_score} - \theta) = -\infty$ \;

    \hfill \\
    \tcp{Update non-blank, non-repeated states}
    \If{$\texttt{LM}$ \textbf{or} $\texttt{BT}$}{
        $\texttt{lm\_scores},\,\texttt{lm\_sts} \gets \mathrm{LM}\left(\texttt{lm\_sts},\,\texttt{new\_labels}\right)$\;        $\texttt{pb\_scores},\,\texttt{pb\_sts}\gets\mathrm{PB}\left(\texttt{pb\_sts},\texttt{new\_labels}\right)$\;
    }

    \hfill \\
    \tcp{Finally, store new values}
    $\texttt{beams}.\mathrm{update}\left(\texttt{new\_beamid}, \texttt{new\_labels}, \right.$ \\ \hspace{3.4cm} $\left. \texttt{acc\_scores}, \texttt{active\_mask}\right)$\; 
    $\mathrm{RecombineHypotheses}\left(\texttt{beams}\right)$ \;
}
\If{$\texttt{LM}$}{
    \tcp{Update with EOS scores from LM}
    $\texttt{beams.scores} \mathrel{+}= \alpha_{LM} \cdot \mathrm{LM.Final}\left(\texttt{lm\_sts}\right)$
}

\label{algorihtm:ctc_beam_search}
\end{algorithm}


\subsection{Batched Beam search algorithm}
Our implementation is inspired by the CTC beam search algorithm from the Flashlight framework, which was originally designed for single-CPU execution. We restructured the algorithm to rely exclusively on vectorized operations, making it more suitable for efficient GPU execution. Unlike the original Flashlight implementation, which processes input samples sequentially within a batch and loops over beam hypotheses at each time step, our approach leverages both input-level and intra-input parallelization. By processing the entire batch simultaneously and vectorizing operations across hypotheses, we eliminate separate iterations over individual inputs and hypotheses -- leaving only a single loop over time steps. 

A key feature of CTC models is that each output label corresponds to a specific encoder time step, producing one label (or blank) per frame. The final transcript is formed by merging repeated labels and removing blanks. Since CTC is non-autoregressive, the model computes log-probabilities for all encoder frames simultaneously, resulting in an output tensor of shape \((1, T, V)\), where \(T\) is the number of time steps and \(V\) is the vocabulary size (including the blank token).

When processing batches, shorter sequences are padded to match the longest one. Then, output log probabilities is tensor $D \in \mathbb{R}^{B \times T \times |V|}$, where \(B\) is the batch size and \(T\) is the maximum number of time steps in the batch. A vector of valid lengths $L \in \mathbb{N}^{B}$ is also returned to indicate the true length of each sequence before padding.

The algorithm takes as input a $D$ and $L$ and returns a \texttt{BatchedBeamHyps} structure containing the decoded transcripts. High-level overview of the algorithm is shown in Algorithm \ref{algorihtm:ctc_beam_search}. 

Algorithm starts by initializing initial scores per batch. At the beginning of the algorithm, only the first hypotheses in each beam are active, all others are inactive with score set to $-\infty$ to ensure a proper starting point. LM and boosting tree scores (BT) are initialized differently: for LM scores for start-of-sentence tokens are retrieved, and for boosting tree scores are obtained for the initial state.

The decoding proceeds time step by time step. For each $t$, to ensure correct processing, active hypotheses mask is computed using current time step index and valid length of specific sample. The core update computes candidate token scores by combining model log probabilities with the accumulated beam scores. Then word insertion penalty $\beta$ is added unless the token is a blank or repetition. When active, LM and BT scores are added to non-blank and non-repeating tokens, weighted by $\alpha_{\text{LM}}$ and $\alpha_{\text{BT}}$, respectively. We also experimented with scoring repeated token emissions using the LM or BT at each occurrence, rather than only on their first appearance, but observed no improvements. 

Next, the decoder performs a flat \texttt{TopK} selection across all beams and labels. Hypotheses below a threshold $\theta$ from the top score are pruned. If LM or BT is active, their internal states are updated for all non-blank and non-repeating labels.

Finally, beam states are updated with new scores and labels, and recombination merges identical hypotheses. At the final step, if LM is used, its end-of-sequence scores are added to the beam scores.

\subsection{CUDA Graphs}

During beam decoding, the GPU performs many small, repeated operations at each step. With small beam and batch sizes, each kernel performs minimal computation, making the kernel launch overhead a significant bottleneck, often dominating the actual execution time. To address this, we use CUDA Graphs to capture the decoding workload into a static execution graph that can be replayed. This reduces kernel launch overhead and minimizes CPU-GPU synchronization delays.
 
\section{Experimental setup}

\begin{table}[ht]
\caption{Comparison of FlexCTC and greedy decoding customizations. \textbf{Batch size:} 32, \textbf{Beam size:} 8.}
\centering
\begin{tabular}{l@{\hspace{6pt}}l|c@{\hspace{6pt}}c@{\hspace{6pt}}c|c@{\hspace{6pt}}c@{\hspace{6pt}}c}
\toprule
 & \multirow{2}{*}{\textbf{Mode}} & \multicolumn{3}{c|}{\textbf{MultiMed}} & \multicolumn{3}{c}{\textbf{Earnings21}} \\
 &  & \textbf{WER$\downarrow$} & \textbf{Fscore$\uparrow$} & \textbf{RTFx$\uparrow$} & \textbf{WER$\downarrow$} & \textbf{Fscore$\uparrow$} & \textbf{RTFx$\uparrow$} \\
\midrule
\multirow{4}{*}{\rotatebox{90}{greedy}} & no    & 15.09 & 55.5 & 2804 & 15.02 & 69.3 & 2687 \\
                                        & PB    & 14.96 & 60.7 & 2685 & 14.97 & 72.0 & 2521 \\
                                        & LM    & 14.89 & 58.0 & 2626 & 14.76 & 70.0 & 2477 \\
                                        & LM+PB & 14.83 & 60.8 & 2573 & 14.73 & 72.2 & 2445 \\
\midrule
\multirow{4}{*}{\rotatebox{90}{beam}}   & no    & 15.09 & 55.5 & 2306 & 15.02 & 69.2 & 2096 \\
                                        & PB    & 14.47 & 70.2 & 2106 & 14.76 & 77.8 & 2013 \\
                                        & LM    & 14.00 & 66.1 & 2075 & 13.92 & 72.1 & 1978 \\
                                        & LM+PB & \textbf{13.55} & \textbf{74.2} & 1995 & \textbf{13.74} & \textbf{79.6} & 1885 \\
\bottomrule
\end{tabular}
\label{tab:custom}
\end{table}

\subsection{Models}

In our experiments, we use the Fast Conformer CTC Large model~\cite{model:ctc}, which has approximately 115 million parameters and is trained on 24k hours of diverse English speech. The model employs a byte-pair encoding (BPE) tokenizer with a vocabulary size of 1,024.

For LM fusion in our method and Flashlight, we use 6-gram subword-level language models built with the KenLM library. NGPU-LM models are constructed from the same \texttt{ARPA} file to ensure consistency between Flashlight and our approach. For CUDA WFST and PyCTCDecode---which require word-level n-gram models---we build separate 4-gram language models, assuming an average of $\sim$1.5 tokens per word. SPGI language models are pruned with a threshold of 1.

\begin{table*}[ht]
\centering
\caption{Comparison with other methods in LM fusion. \textbf{Batch size:} 32, \textbf{Beam size:} 8}
\resizebox{0.8\textwidth}{!}{
\begin{tabular}{l|c|cc|cc|cc}
\toprule
\multirow{2}{*}{\textbf{Strategy}} & \multirow{2}{*}{\textbf{Params}} 
& \multicolumn{2}{c|}{\textbf{SPGI}} 
& \multicolumn{2}{c|}{\textbf{Earnings21}} 
& \multicolumn{2}{c}{\textbf{MultiMed}} \\
& & \textbf{WER$\downarrow$} & \textbf{RTFx$\uparrow$} & \textbf{WER$\downarrow$} & \textbf{RTFx$\uparrow$} & \textbf{WER$\downarrow$} & \textbf{RTFx$\uparrow$} \\
\midrule
Greedy       &      no LM      & 6.41          & 2797    & 15.02 & 2687 & 15.09 & 2804 \\
\midrule
PyCTCDecode  & beam = 4        & 4.69          & 1210    & 14.37 & 1129 & 15.45 & 1188 \\
Cuda WFST    & max = 1k        & 4.64          & 1065    & 14.33 & 419  & 16.77 & 778 \\
Flashlight   & beam = 4        & \textbf{4.48} & 1071    & 14.02 & 1100 & 14.17 & 1279 \\
\textbf{Our: FlexCTC} & beam = 4  & \textbf{4.48} & \textbf{2085} & \textbf{13.98} & \textbf{2084} & \textbf{14.16} & \textbf{2212} \\
\midrule
PyCTCDecode  & beam = 16       & 4.50          & 909     & 14.23 & 731  & 14.99 & 751 \\
Cuda WFST    & max = 10k       & 4.40          & 832     & 14.68 & 782  & 16.56 & 395 \\
Flashlight   & beam = 16       & 4.39          & 454     & \textbf{13.89} & 686  & \textbf{13.93} & 631 \\
\textbf{Our: FlexCTC} & beam = 16 & \textbf{4.38} & \textbf{1928} & \textbf{13.89} & \textbf{1835} & \textbf{13.93} & \textbf{1956} \\
\bottomrule
\end{tabular}
}
\label{tab:strategies}
\end{table*}
\subsection{Datasets}
We conduct our experiments with LM fusion and phrase-boosting on three publicly available datasets, representing the financial and medical domains, not explicitly observed during training. 

\textbf{SPGI} dataset~\cite{article:spgi} is a large-scale collection of transcribed earnings calls from publicly traded U.S. companies, comprising 5,000 hours of professionally recorded speech in the financial domain. Since it does not include a list of domain-specific terms, we do not perform phrase-boosting experiments on this dataset. \textbf{Earnings21} dataset~\cite{article:earnings21} is a benchmark collection of financial earnings calls. It includes approximately 39 hours of long-form audio recordings and provides a domain-specific word list to support phrase boosting. For our evaluations, we segmented the original long-form audio into sentences with a maximum duration of 70 seconds. For both Earnings21 and SPGI LM fusion experiments we built n-gram LM on SPGI train text.

\textbf{MultiMed:} We use the English subset of the MultiMed dataset~\cite{article:multimed}, comprising approximately 77 hours of training data in the medical domain, for LM fusion and phrase boosting experiments. The ``test'' subset serves as the test set, and the ``eval'' subset is used as development set. During preprocessing, we observed a mismatch between audio recordings and their transcriptions in both subsets. To address this, we re-transcribed the audio using two base models and filtered out samples where the word error rate (WER) between the original transcription and the new one exceeded 30\%.

MultiMed transcripts contain a large number of specialized medical terms, making the dataset well-suited for phrase-boosting experiments. We extracted a domain-specific word list from MultiMed to support these experiments. The LLaMA 3.3-70B Instruct model~\cite{model:llama3_70b_instruct} was prompted to tag medical terms-such as diseases, procedures, medications, and clinical conditions-within the text. To identify personal names, brand names, and organizations, we used the SpaCy toolkit~\cite{article:spacy}. We filtered the extracted phrases to exclude common or easily recognized words. Specifically, we removed words that were always correctly recognized, appeared more than 10 times set with over 70\% accuracy, occurred fewer than twice, or were shorter than three letters. This filtering resulted in a list of approximately 1,000 medical terms used in our phrase boosting experiments. For LM fusion experiments we used LM constructed on training set transcripts. Numbers are reported after normalization from \cite{article:open_asr_leaderboard}.


\subsection{Metrics}

We evaluate decoding accuracy using WER and F-score. While WER measures overall transcription quality, the F-score is used specifically to evaluate phrase-boosting effectiveness. To assess inference speed, we report inverse Real-Time Factors (RTFx), computed after a single warm-up run and averaged over three runs. All performance measurements are conducted on a single NVIDIA A5000 GPU using \texttt{float32} arithmetic and Intel(R) Core(TM) i9-10940X CPU @ 3.30GHz.

\subsection{Setups}
We compare our method against two CPU-based decoders; Flashlight and PyCTCDecode, and GPU-based Cuda WFST decoder. All three decoders are available within the NeMo framework~\cite{article:nemo}. For all methods, we set the pruning threshold $\theta$ to $12$. Due to its architecture, the WFST decoder uses a different set of parameters; instead of a beam size, it limits the number of active hypotheses via a \emph{maximum active} threshold, making direct comparison challenging. Both decoding accuracy and speed are highly sensitive to this parameter: higher values improve accuracy but reduce speed. For a fair comparison, we selected parameter values that yield similar accuracy in comparable speed.

\begin{table}[t]
\centering
\caption{Comparison with PyCTCDecode in phrase-boosting for 100 boosted words. \textbf{Batch size:} 32, \textbf{Beam size:} 8, \textbf{Test set:} Earnings21}
\label{tab:boosting100}
\resizebox{0.48\textwidth}{!}{
\begin{tabular}{lccc}
\toprule
\textbf{Method} & \textbf{WER $\downarrow$} & \textbf{F-score 100 $\uparrow$} & \textbf{RTFx $\uparrow$} \\
\midrule
Greedy                  & 15.02 & 73.2 & 2687 \\
PyCTCDecode (100)       & 15.16 & 76.7 & 28 \\
Ours (100)              & \textbf{14.93} & \textbf{78.2} & 2068 \\
\bottomrule
\end{tabular}
}
\end{table}

\section{Results}
\subsection{FlexCTC}
Table~\ref{tab:custom} compares greedy decoding and beam search decoding under different customization settings. While greedy decoding achieves the highest decoding speed, it is unable to fully leverage the improvements offered by the customizations, providing only modest relative WER reductions of approximately 1.7\% on MultiMed and 1.9\% on Earnings21, with F-scores increasing by 5.3 points on MultiMed and 2.9 points on Earnings21. In contrast, beam search decoding, although slower, consistently outperforms greedy decoding in accuracy. With the combined LM+PB customization, it achieves the best results, reducing WER by approximately 10.5\% relative on MultiMed and 8.5\% relative on Earnings21. Similarly, the F-score improves substantially with beam search and LM+PB customization, increasing by 18.7 points on MultiMed and 10.4 points on Earnings21. This accuracy gain comes at an approximate 19.9\% relative slowdown compared to greedy decoding without customizations, and about a 22.4\% slowdown compared to greedy decoding with LM+PB. Analysis of the individual contributions shows that phrase boosting primarily enhances phrase-level recognition, resulting in improved F-scores, while language model fusion more broadly reduces WER by improving overall recognition accuracy. Importantly, these two methods do not interfere with each other, and their complementary effects in the combined LM+PB configuration deliver the best trade-off between accuracy and speed.

\begin{figure}[t]
    \centering
    \includegraphics[width=0.48\textwidth]{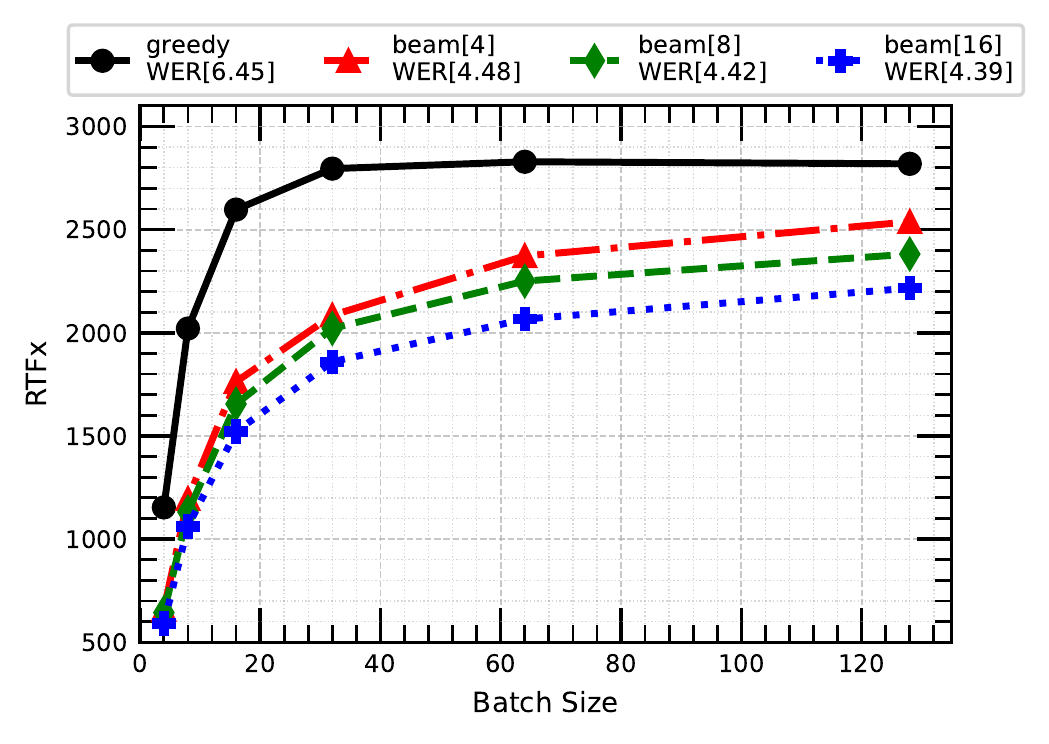}
    \caption{RTFx vs batch size for greedy and beam (LM fusion is enabled) decoding. \textbf{Beam sizes:} 4, 8, and 16. \textbf{Test set: }SPGI}
    \label{fig:batch_size_vs_rtfx}
\end{figure}

\subsection{Comparison with other methods}

Table \ref{tab:strategies} compares FlexCTC’s decoding strategies with several methods supporting LM fusion. Our FlexCTC toolkit consistently achieves the best balance of accuracy and decoding speed. It matches or slightly surpasses the best reported WERs while delivering 2 to 3 times faster decoding performance.

FlexCTC is also is more efficient at handling of larger beam sizes, where other methods suffer significant slowdowns. For example, Flashlight’s decoding speed on MultiMed drops by approximately 2 times. In contrast, FlexCTC experiences only an 11.6\% relative slowdown. This efficiency is due to FlexCTC’s vectorized processing of beam hypotheses, which allows it to scale more gracefully with larger beam widths.

PyCTCDecode and CUDA WFST rely on word-level language models, which tend to perform poorly when training data is limited. This limitation is evident on the MultiMed dataset, where both methods yield significantly higher WER. In contrast, FlexCTC supports subword-level language models, making it more robust in low-resource settings.

Table~\ref{tab:boosting100} compares our toolkit with PyCTCDecode on phrase boosting using a subset of 100 boosted words, extracted from a larger list of about 1000 words. PyCTCDecode’s decoding speed degrades severely with large boosting lists, necessitating this subset for practical evaluation.

Our approach achieves better accuracy, with a WER of 14.93\% versus 15.16\% for PyCTCDecode, and a higher F-score. In addition, our toolkit is approximately 74 times faster than PyCTCDecode on the 100-word subset. For reference, decoding with the full boosting list ($\sim$1000 words) in FlexCTC results in an RTFx of around 2013 (see Table \ref{tab:custom}), indicating minimal speed degradation with larger phrase lists. These results demonstrate that our method delivers substantial improvements in both decoding speed and accuracy when applying phrase boosting to large vocabularies.

\subsection{Scalability}
\textbf{Increasing batch size:} Figure~\ref{fig:batch_size_vs_rtfx} shows how decoding speed changes with batch size for greedy decoding and beam search at different beam sizes. As batch size increases, beam search becomes closer in speed to greedy decoding. At batch size 128, the gap is only 10.3\% for beam size 4 and 21.4\% for beam size 16.

\begin{figure}[t]
    \centering
    \includegraphics[width=0.48\textwidth]{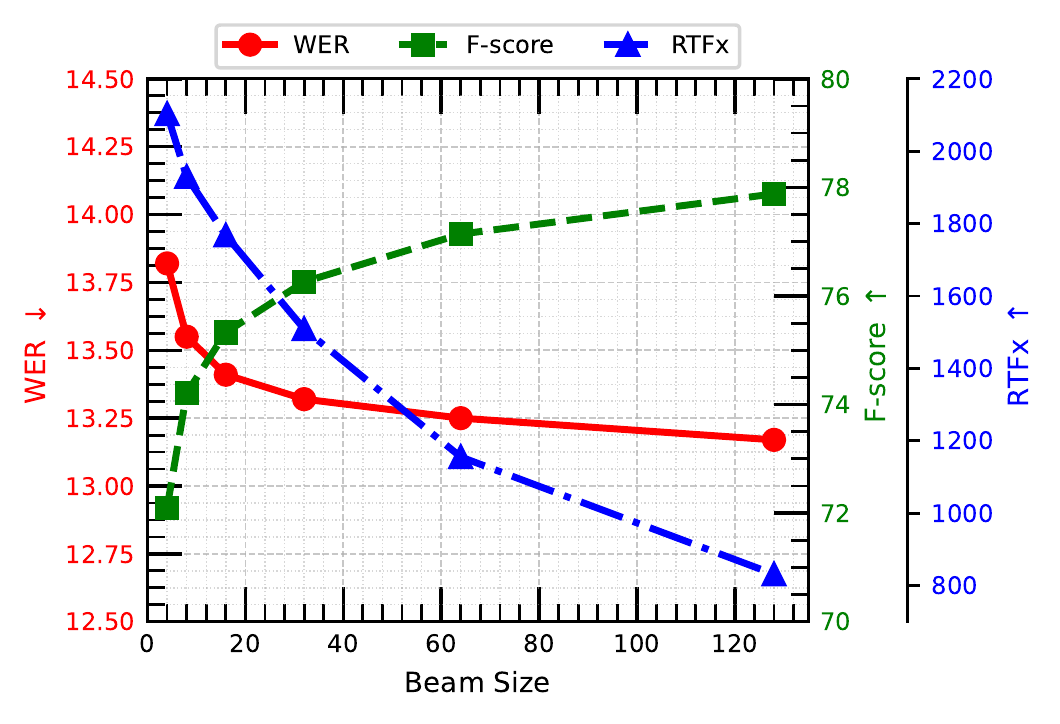}
    \caption{WER(red), F-score(green), and RTFx(blue) vs beam size. LM fusion and phrase boosting are enabled. \textbf{Batch size: }32. \textbf{Test set:} MultiMed}
    \label{fig:beam_size_vs_wer}
\end{figure}

\textbf{Increasing beam size:} Our fully batched decoder design enables efficient expansion to large beam sizes, supporting beam sizes of 128 and beyond, without overhead from sequential processing of hypotheses in beam. In Fig.~\ref{fig:beam_size_vs_wer}, the dependency of WER, F-score, and RTFx on beam size is shown, with beam sizes up to 128. As the beam size increases, recognition quality improves steadily. The WER decreases from 13.82\% at beam size 4 to 13.17\% at beam size 128, resulting in additional absolute WER reduction of 0.65\%, and total 12.7\% relative WER reduction compared to greedy. Additionally, the contextual F-score improves from 72.1\% to 77.9\%, achieving an additional absolute gain of 5.8\%, and total 22.4\% compared to greedy (see. Table \ref{tab:custom}). These improvements highlight the decoder’s increased ability to capture relevant hypotheses under broader search, leveraging language model fusion and word boosting potential. Although RTFx decreases by approximately 2.5 times at beam size 128, it remains at a reasonable level, making the decoder still suitable for practical use cases where accuracy is critical.

\section{Conclusion}

This work introduces a fully GPU-accelerated CTC beam decoding toolkit that bridges the gap between research flexibility and production-ready performance. By integrating a GPU-optimized, batched beam decoder with CUDA Graphs, our framework achieves 2-3 times speedup in inverse real-time factor (RTFx) on benchmark datasets, while maintaining the best accuracy. The toolkit enables efficient contextualization through N-gram LM shallow fusion and phrase boosting, delivering accuracy improvements with minimal latency overhead.

Unlike existing C++/CUDA or WFST-based decoders, our Python/PyTorch-native implementation ensures seamless integration with modern deep learning workflows, lowering adoption barriers for both researchers and engineers. The open-source release under a permissive license further democratizes access to high-performance CTC decoding, accelerating innovation in speech recognition and sequence modeling.

\bibliographystyle{IEEEtran}
\newpage
\bibliography{mybib}

\end{document}